\documentclass[12pt]{iopart}

\usepackage{iopams}
\usepackage{graphicx}
\begin{document}

\title[Nonequilibrium steady states in a vibrated-rod
monolayer]{Nonequilibrium steady states in a vibrated-rod
monolayer : tetratic, nematic, and smectic correlations}

\author{Vijay Narayan$^1$, Narayanan Menon$^{1,2}$ and Sriram Ramaswamy$^1$}
\ead{vj@physics.iisc.ernet.in}
\address{$^1$ Centre for Condensed Matter Theory, Department of
Physics\\ Indian Institute of Science, Bangalore, India
}%
\address{$^2$ Department of Physics, University of
Massachusetts, Amherst, U.S.A.}%

\begin{abstract}
We study experimentally the nonequilibrium phase behaviour of a
horizontal monolayer of macroscopic rods. The motion of the rods
in two dimensions is driven by vibrations in the vertical
direction. Aside from the control variables of packing fraction
and aspect ratio that are typically explored in molecular liquid
crystalline systems, due to the macroscopic size of the particles
we are also able to investigate the effect of the precise shape of
the particle on the steady states of this driven system. We find
that the shape plays an important role in determining the nature
of the orientational ordering at high packing fraction.
Cylindrical particles show substantial tetratic correlations over
a range of aspect ratios where spherocylinders  have previously
been shown \cite{Bates_Frenkel} to undergo transitions between
isotropic and nematic phases. Particles that are thinner at the
ends (rolling pins or bails) show nematic ordering over the same
range of aspect ratios, with a well-established nematic phase at
large aspect ratio and a defect-ridden nematic state with
large-scale swirling motion at small aspect ratios. Finally,
long-grain, basmati rice, whose geometry is intermediate between
the two shapes above, shows phases with strong indications of
smectic order.

\end{abstract}

\pacs{61.30.Eb,45.70.-n}
\maketitle

\section{\label{Intro}Introduction}

Our experiments study the orientational ordering of horizontal,
quasi-two-dimensional assemblies of millimeter-sized, elongated
particles driven into motion by external vibration in the vertical
direction. Since the particles we study are rod-like and do not
deform appreciably, the natural point of comparison is the phase
behaviour of thermal equilibrium hard-rod systems in two
dimensions. The experiments address some basic questions on how
systems in driven, nonequilibrium steady states differ in their
phase behaviour and dynamics from similar systems at thermal
equilibrium: Can the nonequilibrium phase diagram be rationalized
by a comparison to hard-rod Monte Carlo simulations
\cite{Bates_Frenkel} alone or must we invoke new physics? In
thermal equilibrium, the ordering of hard-rod liquid crystals is
determined solely by geometrical considerations of packing, which
should apply equally in the molecular and macroscopic realms.  On
the other hand, thermal fluctuations are of no importance in the
systems we report in this article, and all exploration of phase
space is made possible only by means of external driving forces.

The nonthermal nature of the system has several implications:
well-established extremal principles involving free energy and
entropy cannot be invoked, the sources of fluctuation and
dissipation are distinct, and nonequilibrium fluctuations may
either help or hinder certain kinds of order. `Moving XY models'
of flocking \cite{Vicsek,tonertu}, for instance, have been shown
to display true long-range order in two dimensions, unlike their
equilibrium counterparts. Nonequilibrium steady states with
strictly nematic order \cite{sradititoner} are predicted to have
anomalously large density fluctuations, and velocity
autocorrelations that decay much slower than in corresponding
thermal systems. There are also a number of papers
\cite{Kardar,Liverpool_Marchetti,Gautam_Menon,Tsimring_Aranson} on
the dynamics of orientable systems that are motivated by the
biologically important motor-microtubule system. These systems
support a variety of nonequilibrium structures, in particular,
rotating vortices, which have no equilibrium analogue. While the
particles in these systems are polar in nature and therefore not
directly comparable to our particles, they do show the
possibilities of qualitatively new phases and defect behaviours in
systems of self-driven orientable particles \cite{tonertusr, AransonTsimringReview}.

In the context of nonthermal systems like granular media, Edwards
and coworkers have taken an entirely different theoretical
approach: they propose a statistical mechanics for granular
systems by constructing an ensemble that includes with equal
probability all mechanically stable realizations of a system. The
phase behaviour of a system of anisotropic grains has been
considered \cite{MounfieldEdwards} within this framework. The
theory predicts, remarkably, that elongated particles do not
undergo an isotropic-to-nematic transition as a function of
density.

There have also been some recent experiments and simulations on
the packing of anisotropic granular particles under gravity. Monte
Carlo \cite{BuchalterBradley92} simulations of ellipses falling in
2-dimensions found that they tend to orient with their long axes
perpendicular to gravity. Quasi-2D experiments and simulations
\cite{StokelyFranklin03} on cylinders showed similar results, with
nematic ordering extending over two particle lengths. Villarruel
{\it et al.} \cite{Villarruel} studied the compaction of rods in
3-dimensions: rods were poured into a cylindrical container and
compacted by 'tapping' i.e. applying isolated vertical jolts.
While quantitative measures of orientational order were not
presented, it was observed that repeated tapping caused the rods
to orient vertically, first at the walls and then in bulk, into
smectic-like layers. Blair {\it et al.} \cite{blair} have studied
vibrated systems of rods, contained in circular and annular
geometries. At high packing fractions, the rods separated into
regions where they are stacked horizontally, and regions in which
rods came out of the horizontal plane and are only slightly tipped
from the vertical; in the latter regions, vortices formed,
rotating in a direction governed by the tilt of the rods. A
phenomenological explanation for the formation and motions of the
vortices was proposed in Ref. \cite{AransonTsimring03}. Although
we do report on nonequilibrium dynamical effects such as these,
our major preoccupation here is with static orientational
ordering. Also, the role of gravity in our experiments is to be
distinguished from those reported in Refs.
\cite{BuchalterBradley92, StokelyFranklin03, Villarruel}: while
gravity is very important in the kinematics of the particle motion
it does not play the role of an ordering field as in
\cite{BuchalterBradley92, StokelyFranklin03, Villarruel}. The
ordering in our system arises purely from interparticle
interactions. Measurements of the isotropic-nematic transition in a 
quasi-2D system of vibrated rods have also been reported by \cite{losert}.
The focus in that paper is on understanding the long-wavelength 
distortions of the director in terms of competition between wall alignment
and bulk elasticity.Experiments very similar in intent to ours have been
performed by Fraden and coworkers \cite{Fraden04}.

While there are no firmly-founded theoretical predictions to guide
the effort of charting out a nonequilibrium phase diagram for
elongated particles, simulations of hard rods in equilibrium
provide a valuable point of reference. Frenkel and others
\cite{Bates_Frenkel,Lagomarsino} have performed Monte Carlo
simulations in two dimensions, on spherocylinders of length $L$
and diameter $D$. They find that nematic phases do not form for
particle aspect ratio ${\cal A}\equiv L/D < 7$. Instead, they
identify an isotropic phase at low number densities and a layered
solid at high densities. For particles with ${\cal A}>7$ a nematic
phase intervenes at intermediate area fractions. In a grand canonical
Monte Carlo study in two dimensions, Khandkar and Barma 
\cite{khandkarbarma} confirm the quasi-long-range nature of 
nematic ordering in the ``needle'' (i.e., $D = 0$) limit, 
and demonstrate that imposed gradients in the chemical potential
suppress director fluctuations. A recent Monte Carlo simulation 
\cite{DonevTorquato05} studies the ordering of rectangles with 
${\cal A}=2$ and finds a transition from isotropic to tetratic 
order. In an earlier work, Veerman {\it et al.} \cite{Veerman}
report the occurrence of a cubatic phase in a system of truncated 
spheres and more recently Borukhov {\it et al.} \cite{Borukhov} 
find cubatics in linker-filament models for the cytoskeleton. While
we are able to make a detailed comparison to these equilibrium 
simulations, due to the macroscopic nature of the 
particles we are also able to explore a new axis in the study of 
hard-rod systems: we can make essentially arbitrary variations 
in the shape of the particles by selectively etching cylindrical 
rods. We find a surprising dependence on the precise shape of 
the particle -- objects with similar aspect ratios but dissimilar 
shapes order differently. It is not possible to determine on 
the basis of our work alone whether this shape-sensitivity is 
a nonequilibrium effect, however, we hope to stimulate more 
detailed investigations of these effects in hard-particle simulations.

Following this introductory section is an experimental section,
\ref{Exp} in which we describe the particles used in our study,
the geometry of the cell, and the driving mechanism. The system is
visualized by real-space images; the acquisition and analysis of
the images is also outlined.  A detailed  description of our
results is in section \ref{Results}, but we summarize these
briefly here: Strictly cylindrical particles show substantial
tetratic (that is, four-fold orientational) correlations at both
small (${\cal A}= 4.9$) and large (${\cal A}= 12.6$) aspect ratio.
The length scale of the correlations increases as the number
density is increased; however, we do not see long-range order
(LRO) over the range of area fractions and vibration amplitudes we
explore.  Particles that are thinner at their tips (``rolling
pins'' or ``bails'', see figure \ref{Particle_types}) show nematic 
ordering at high area fractions
in the same range of aspect ratios. For shorter particles (${\cal
A}= 5.2$) we obtain a defect-ridden nematic state with large-
scale swirling motion at the number densities we study in our
experiment. For ${\cal A}= 10.4$, we find at high densities, a
phase with at least quasi-long-range nematic order (QLRO), that
is, power-law decay of nematic correlations.  We can induce
transitions out of this nematic phase either by lowering number
density or by increasing the intensity of the driving. Finally,
long-grain, basmati rice, with an aspect ratio of ${\cal A}\approx
4$, shows phases with strong indications of smectic order. We
comment on this below in section \ref{smectics}. We also find
intriguing dynamical phenomena, with clear signatures of
nonequilibrium behaviour, as discussed at the end of section
\ref{Results}.

\section{\label{Exp}Experimental setup and procedure}

Our experiments involve vertically shaking a horizontal layer of
rod-like particles. The rods were confined to a single layer
between an aluminium substrate and a lid that fit over the
particles leaving them only a little vertical room to move in.  In
this section we describe the particles, the geometry of the
experimental cell, the driving mechanism and, finally, the methods
of image acquisition and analysis.

\subsection{\label{Particles}Particle Dimensions}

\begin{figure}
\begin{center}
\includegraphics[width = 3.25 in]{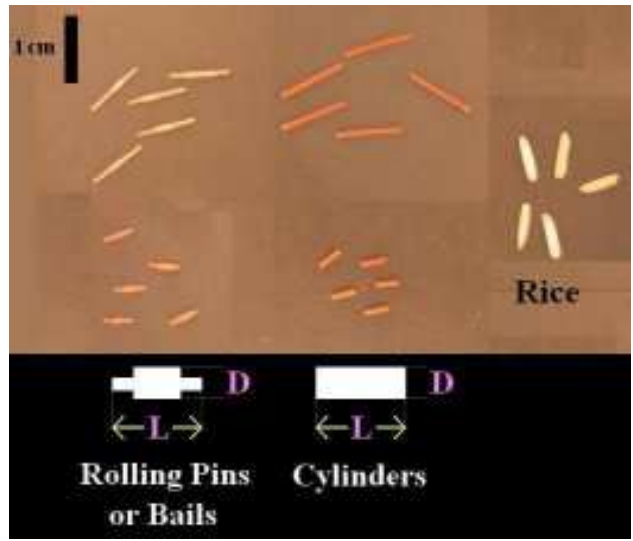}
\caption{\label{Particle_types}The 5 kinds of particles we used in
our experiments. White silhouettes are schematic outlines of the
2D projections of the particles. We call particles with a strictly
rectangular profile `cylinders' and the others `rolling pins' or
`bails'. We define the aspect ratio as ${\cal A}\equiv L/D$ where
the diameter $D$ is measured at the largest cylindrical
cross-section. Far right: basmati rice grains. For details of size
and polydispersity of particles see Table \ref{particlefactfile}}
\end{center}
\end{figure}

\begin{table}
\caption{\label{particlefactfile}Tabulated below are the details
of the various particles we used in our experiments.}
\begin{center}

\begin{tabular}{|c|cc|}
\hline
  & Aspect Ratio & Length (mm) \\
  \hline
  Rice& 4.0 $\pm$ 1.07 & 7.1 $\pm$ 0.44\\
  Short cylinders & 4.9 $\pm$ 0.42 &  3.7 $\pm$ 0.11\\
  Short rolling pins & 5.2 $\pm$ 0.59 &  4.6 $\pm$ 0.16\\
  Long cylinders & 12.6 $\pm$ 0.66 &  9.9 $\pm$ 0.12\\
  Long rolling pins & 10.4 $\pm$ 1.45 &  9.2 $\pm$ 0.39\\
\hline
\end{tabular}
\end{center}
\end{table}

We used five different kinds of particles in the experiment.
Figure \ref{Particle_types} shows images of a few of each. The
particles labelled ``cylinders" are cut from a length of enamelled
copper wire.  The process of cutting introduces only a small
degree of polydispersity (less than 5\%) in the length of the
cylinders. The particles labelled ``rolling pins'' (or ``bails''
\cite{bailfootnote}) are produced by etching the tips of the
cylinders in ferric chloride solution. Since there are variations
in the local etch rate, and in the exact length of rod that is
exposed to the etchant, the degree of polydispersity in the bails
is somewhat greater. The rice was used as purchased. Table
\ref{particlefactfile} summarises the dimensions of the particles
used in the experiment.

\subsection{\label{Cell}Geometry of the Cell}

\begin{figure}
\begin{center}
\includegraphics[width = 3.25 in]{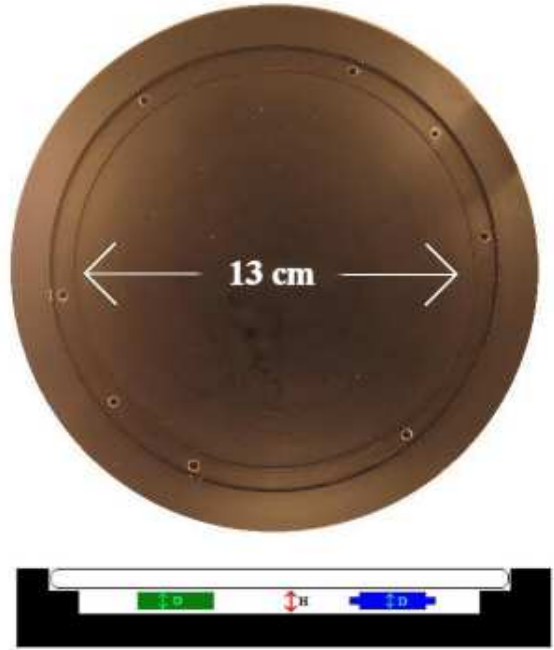}
\caption{\label{plate}Top and side view of base of circular sample
cell. The height of the sample volume, $H=1.0$ mm, and the
diameter of the particles, $D=0.8$ mm.}
\end{center}
\end{figure}

The particles were confined in a shallow circular geometry with a
13 cm diameter, as shown in figure \ref{plate}. Whenever
necessary, it was possible to further confine the particles to a
square region inscribed within the diameter of the cell.  The base
of the cell is made of aluminium, anodized black to enhance
optical contrast with the particles.  The top of the cylinder is
made of perspex, and is placed so as to leave a vertical space $H
= 1$ mm above the base. Since the particles have a diameter $D
\simeq 0.8$ mm, this gives them about 25\% of their thickness to
explore in the vertical direction. Thus particles cannot pass over
one another by moving out of plane. The vertical degree of motion
allows the planar projection of the particles to overlap a little
when they are being shaken; the horizontal component of the
distance of closest possible approach of two particles is slightly
less than $D$. Thus, the vertical clearance $H - D$ can be seen as
a parameter that tunes the particle interaction potential from
hard- core repulsion ($H-D = 0$) when the motion is strictly
2-dimensional, to slightly softened repulsion for $H>D$. The
shaking amplitude $a_{o}$ can then be seen as playing the role of
temperature.

Another important feature of the cell is the boundary condition
presented to the particles at the edge of the system. When the
boundary is smooth, particles tend to lie parallel to the
boundary. This occurs both when the boundary is a square or a
circle since in the latter case the radius of curvature is much
larger than the particle length.  The parallel alignment is a
consequence of dynamical stability: when a rod aligned with the
wall collides with it at one end and rotates out of alignment,
then the next wall-collision will be with the other end of the
rod, which will tend to restore the alignment.  No restoring
torque of this nature exists for perpendicular alignment with the
wall. In some situations, we induce perpendicular alignment by
gluing a layer of rods normal to the wall. The tips of particles
perpendicular to the wall tend to get trapped in the interstices
between the tips of the glued particles. We found this procedure
to be effective only for the bails, and not for the cylinders.

\subsection{\label{Shaking}Excitation Mechanism}

The cells were filled with particles to a known area fraction and
then mounted on a permanent magnet shaker (LDS 406).  For most of
the experiments described here, the cell was shaken at a frequency
$f=200 Hz$ and an amplitude $a_{o}$ resulting in a fixed peak
acceleration, as measured by an accelerometer (PCB Piezotronics
352B02), of $\Gamma \equiv (2 \pi f)^2 a_{o} = 6g$ where $g$ is
the acceleration due to gravity. The corresponding amplitude
typically used was $a_{o}=0.037$ mm = $0.046D$ , smaller than the
clearance of $H-D=0.2D$ between the top of the particle and the
lid. However, the characteristic velocity of the plate ($2\pi
fa_{o}$) is sufficient to launch particles to a height of $\sim
0.1 mm$ . In one instance we varied the acceleration between 5g
and 8g to observe the effect it had on the degree of ordering.

\subsection{\label{Images} Image acquisition and
analysis}

Dynamical features of the steady states were imaged with a
high-speed video camera, however, for the majority of the results
reported in this article, our primary interest was the
instantaneous configurations of the particles. These were imaged
using a consumer digital camera (Canon PowerShot G5) placed above
the cell. Images were taken at a resolution of 2592 $\times$ 1944
pixels, at time intervals of approximately 15 secs. This interval
was long enough for significant rearrangements of the particles to
have taken place, and hence for each image to be a statistically
independent sampling of the steady state arrangement of the
particles. In the rest of this article, we will discuss data
extracted from images that were taken while the sample was being
continuously agitated by the shaker. Images taken by halting the
drive (as was done in \cite{Fraden04}) showed qualitatively
similar behaviour but with noticeably greater disorder in the
particle arrangement.

\begin{figure}
\begin{center}
\includegraphics[width = 3.25 in]{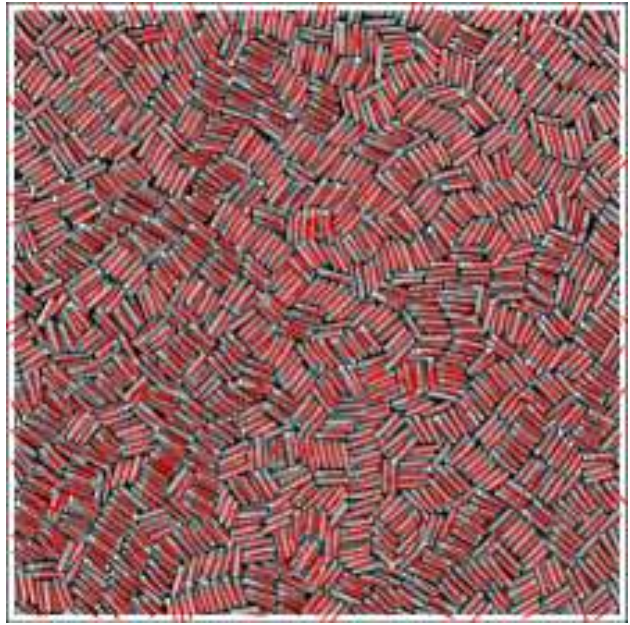}
\caption{\label{particle_id}Particles identified by image analysis
routine: each identified particle was assigned a centre-of-mass
coordinate and an orientation, indicated by the location and
orientation of the bars in the image. Even at these high area
fractions, typically 97\% of particles were identified.}
\end{center}
\end{figure}

The images were analyzed within the ImageJ \cite{ImageJ} analysis
software. The images were subjected to a bandpass filter to remove
pixel noise and long-wavelength variations in illumination, and
then thresholded before identifying the centre of mass and
orientation of each particle. The error in assigning the
coordinates was about 0.5D (it was possible to reduce this error
to 0.2D for the short cylinders) and the error in assigning the
angles was less than $2.5^o$. Figure \ref{particle_id} has bars
indicating the centres of mass and orientation of the particles
superimposed on the image. Typically 97\% of the particles were
successfully identified, the remaining particles being difficult
to resolve as independent objects because of the high packing
density, and the consequent lack of intensity gradient between
particles. The other objects identified are mainly particle pairs;
we are able to determine the position and orientation of some, but
not all, by using information on the minor and major axes of the
composite object. However, due to the relatively small number of
such instances, they did not affect the results we report here.

\section{\label{Results}Results}

\begin{figure*}
\includegraphics[width = 6 in]{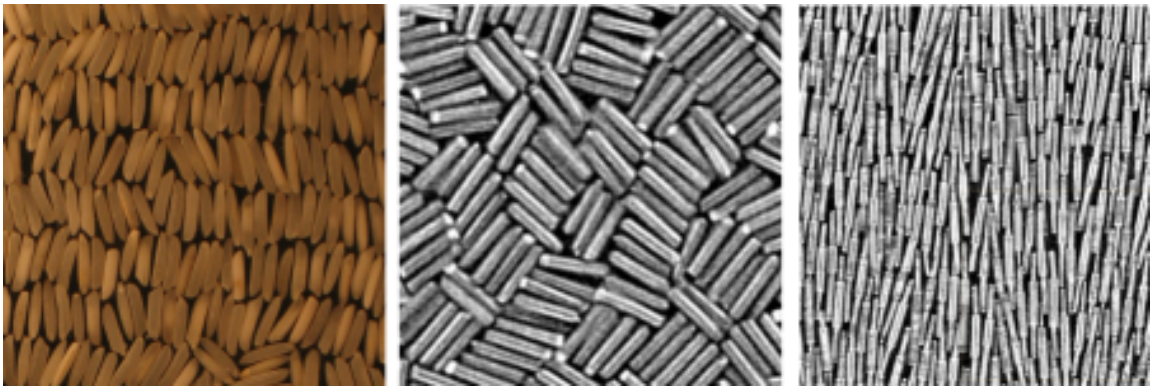}
\caption{\label{realspace}Real space images of steady-states
observed: striped, smectic-like state with the {\em basmati}
grains (left), state formed by aspect ratio ${\cal A}=5.2$
cylinders with strong four-fold correlations (centre) and nematic
state formed by ${\cal A}=10.4$ rolling pins.}
\end{figure*}

This section is organized in the following manner: the first half
will be devoted to results obtained using the cylinders and the
latter half to those with the rolling pins.

\begin{figure}
\begin{center}
\includegraphics[width = 3.25 in]{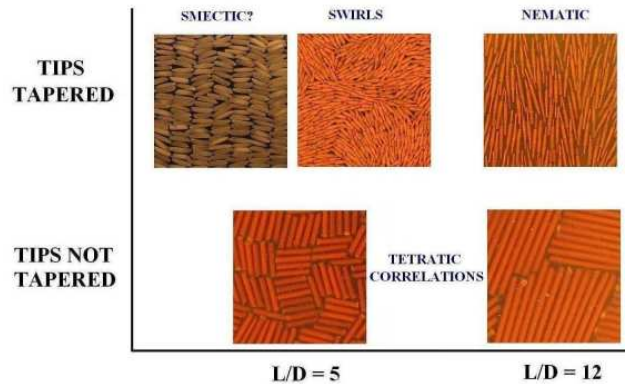}
\caption{\label{phase_diagram}``Phase diagram'' shows region of
parameter space explored and steady states in each region.
Vertical axis represents the two different shapes of the particles
we used.}
\end{center}
\end{figure}

Figure \ref{phase_diagram} shows the region of phase space
explored and gives us an overview of the results. The cylinders,
irrespective of aspect ratio, displayed strong four-fold
orientational, i.e., tetratic, correlations. Studies of the
area-fraction-dependence of these correlations, presented in
subsection \ref{tetratic}, were done using the shorter rods: since
the system size measured in terms of particle length was greater
for the short particles, they were less sensitive to the nature of
the boundary and offered better statistics. We observed
short-ranged tetratic order, but with large correlation lengths.
The high aspect ratio rolling pins formed nematic phases with
quasi-long-range order with power-law decay of orientational
correlations. The low aspect ratio rolling pins formed swirly
states which we discuss in subsection \ref{Swirls}. Common to both
nematic and tetratic states, was the tendency to rotate globally
and persistently in one direction, presumably in response to some
asymmetry in the boundaries of the system. This distinguishes this
phenomenon from that observed by \cite{Tsaietal} who also observe
globally rotating phases, only the particles forming their phase are,
unlike ours, intrinsically chiral. We discuss this in
subsection \ref{Rotation}. We also observed striped, smectic-like
phases in experiments using rice grains. Interestingly, we have
not been able to observe the stripes using bails or cylinders with
the same aspect ratio as the rice.

\subsection{\label{tetratic}Towards the tetratic phase}

Over the range of concentration and other parameters explored, the
strictly cylindrical particles (i.e., flat headed, neither tapered
like the ``bails'' that showed nematic order, nor spherically
capped like the spherocylinders widely studied in simulations) did
not form ordered phases. However, their isotropic phase showed
tetratic, that is, four-fold orientational, short-range order
unaccompanied by nematic correlations. The range of tetratic
correlations increased substantially with increasing area
fraction. The correlation function evaluated was $G_4(r) =
<cos[4{(\theta_i - \theta_j)}]>$, where $\theta_i$ is the
orientation of the rods with respect to some arbitrary,
pre-selected axis. $i$ and $j$ are particle indices that run from
1 to the number of particles, N. The angular brackets are a time
average, over pairs $i,j$ of particles separated by a distance
$r$. For each pair, we calculate $G_4$ as a function of
$r_{\perp}$ and $r_{||}$ which are, respectively, the components
of the separation vector ${\bf r}$ between the rod centres
perpendicular and parallel to the cylinders' long axes.
$r_{\perp}$ and $r_{||}$ are distinct for each particle in a given
pair, they have to be evaluated for each particle, rather than
each particle pair, as would have been sufficient if $G_4$ was
evaluated as a function of the scalar distance, $r$.

\begin{figure}
\begin{center}
\includegraphics[width = 3.25 in]{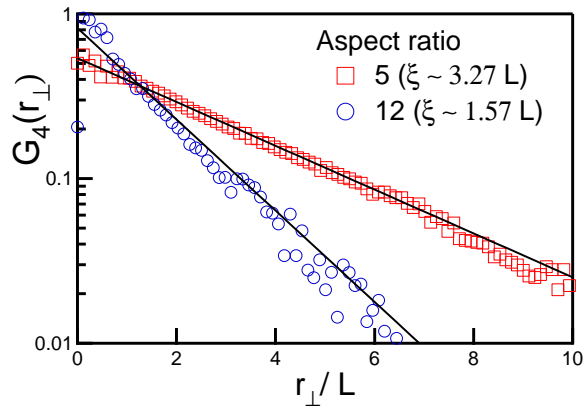}
\caption{\label{SRvsLR} Tetratic correlation lengths (scaled by
rod length $L$) are larger for aspect ratio ${\cal A}=5.2$ than
for ${\cal A}=12.6$. Results shown for area fractions ~87\% and
correlations as function of $r_{\perp}$ (defined in section
\ref{tetratic}); the behaviour with respect to $r_{||}$ is
similar.}
\end{center}
\end{figure}

Both long (${\cal A}=12.6$) and short (${\cal A}=5.2$) cylinders
showed tetratic correlations as shown in figure \ref{SRvsLR}. The
range of tetratic order is shorter (when distance is scaled by
cylinder length) for the long cylinders, however, even a
correlation length $\xi=1.57L=19.8D$, as found for the long
cylinders, indicates substantial tetratic ordering with clearly
identifiable four-fold motifs as seen in Fig.\ref{phase_diagram}.
When the experiments were performed in a square, rather than a
circular frame, one might have expected four-fold orientational
order to be in general enhanced, since alignment parallel to the
walls is (see section \ref{Intro}) favoured by the mechanics of
collisions with the wall. We find that this is true for both long
and short cylinders but not at all for the bails.

\begin{figure}
\begin{center}
\includegraphics[width = 3.25 in]{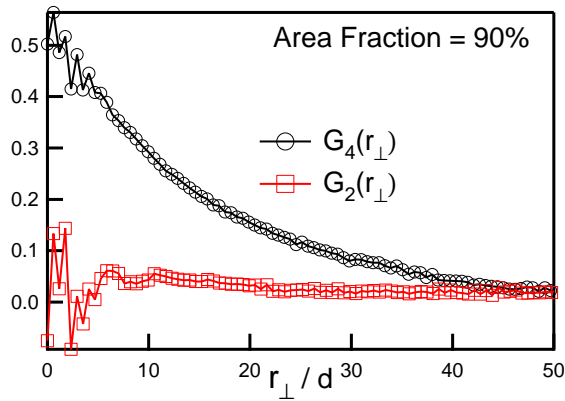}
\caption{\label{G4vsG2}Substantial $G_4(r)$ and very small
$G_2(r)$ for cylinders (${\cal A}=5.2$) at area fraction 90\%
confirm that the correlations are tetratic, not angular harmonics
in two-fold ordered states.}
\end{center}
\end{figure}

To distinguish between tetratic and nematic ordering, we have
plotted for the short cylinders ${\cal A}=5.2$ at area fraction
90\% , $G_4(r)$ and $G_2(r)$ in Fig. \ref{G4vsG2}; here $G_2(r) =
<cos[2{(\theta_i - \theta_j)}]>$ quantifies the extent of nematic
ordering.  $G_4(r)$ clearly dominates, and nematic correlations
are negligible. This confirms what is apparent from the real space
image of the tetratic phase, figure \ref{realspace}. We are thus
seeing true four-fold correlations, not angular harmonics of
2-fold correlations.

\begin{figure}
\begin{center}
\includegraphics[width = 6.5 in]{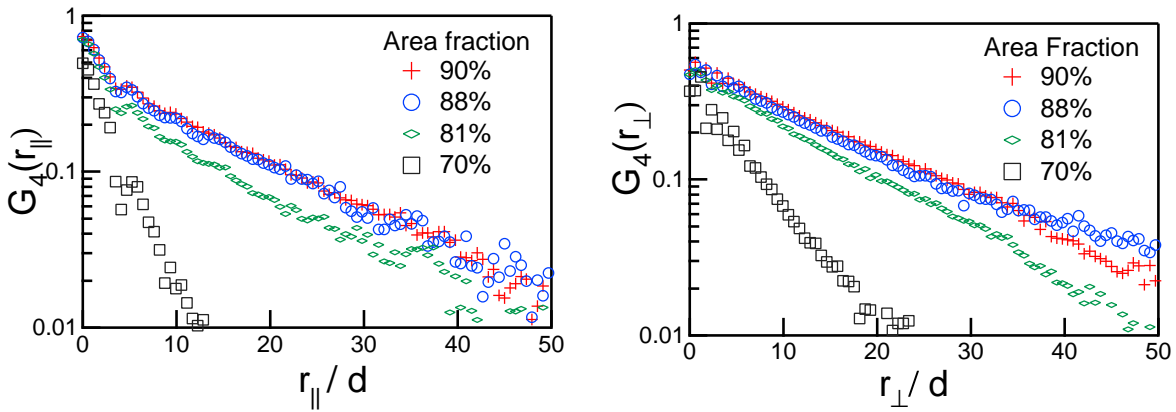}
\caption{\label{G4r} $G_4$ (defined in section \ref{tetratic}) as
a function of $r_{\perp}$ and $r_{||}$ for different area
fractions, for cylinders with aspect ratio 5.}
\end{center}
\end{figure}

\begin{figure}
\begin{center}
\includegraphics[width = 3.25 in]{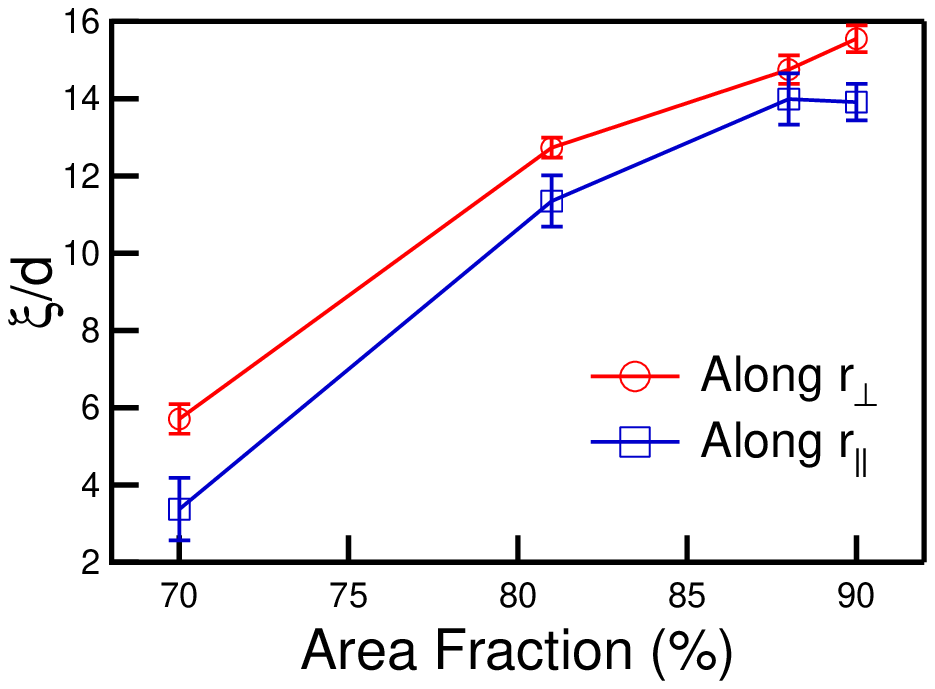}
\caption{\label{corrlength_AF}Tetratic correlation lengths
increase with area fraction, suggesting that a tetratic phase
should appear if some other parameter such as shaking amplitude is
varied. We do not know the origin of the slight difference between
the correlation lengths obtained along $r_{\perp}$ and $r_{||}$.}
\end{center}
\end{figure}

We now present the dependence of the range of tetratic
correlations on area fraction for the short cylinders (${\cal
A}=5.2$). Fig. \ref{G4r} shows $G_4$ as a function of $r_{||}$ and
$r_{\perp}$ for various area fractions. The range of tetratic
correlations increases monotonically with area fraction, as shown
in Fig. \ref{corrlength_AF} which displays the correlation length,
$\xi$ , extracted from Fig. \ref{G4r}. $\xi$ obtained from
$r_{\perp}$ and $r_{||}$ show the same qualitative behaviour, but
are slightly different in magnitude.

The correlation lengths are substantial at high area fractions,
indicating a high tetratic susceptibility, and the likelihood that
a quasi-long-ranged tetratic phase lurks nearby in parameter
space. However, it remains to be determined which parameter --
area fraction, confinement thickness, vibration amplitude or
frequency -- will most easily take the system across this
conjectured isotropic-tetratic phase boundary. We reiterate that
the tetratic ordering is a strong function of particle shape; by
way of comparison, the equilibrium phase diagram of
\cite{Bates_Frenkel} for discorectangles at similar aspect ratios
shows no tendency to tetratic order. They find isotropic, nematic,
and crystalline phases, none of which we see with rectangles. This
difference could be a nonequilibrium effect rather than a shape
effect, however as we will see in section \ref{Nematic}, tapered
particles order differently. Also, Monte Carlo simulations of
rectangles \cite{DonevTorquato05} at much smaller aspect ratio,
${\cal A}=2$, reveal a true tetratic phase. While it is not clear
at all whether this equilibrium phase will survive to the large
${\cal A}$ regime, we hope that our results will stimulate
equilibrium simulations of this region of parameter space.

\subsection{\label{Nematic}The nematic phase}

Why do cylinders not show nematic order even when they are
extremely long?  Nematic order is typically investigated in
simulations using spherocylinders, discorectangles or ellipsoids,
all of which have narrower tips than waists and can thus
interdigitate, promoting uniaxial alignment. This indicates that
the rod tips needed to be tapered to obtain a nematic phase. We
therefore used the particles labelled ``bails'' in figure
\ref{Particle_types}, with aspect ratio ${\cal A}=10.4$, since
equilibrium Monte Carlo studies tell us that ${\cal A} > 7$ is
needed to produce a two-dimensional nematic.

\begin{figure}
\begin{center}
\includegraphics[width = 6.5 in]{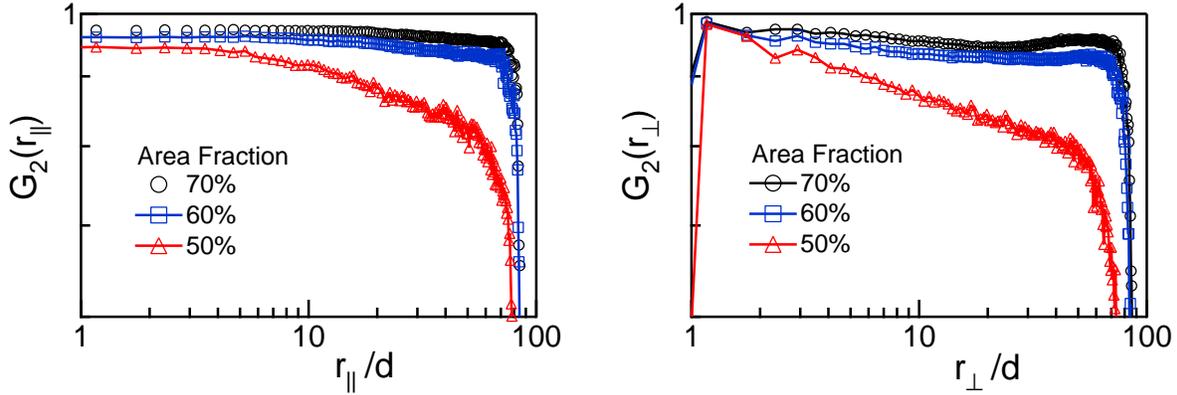}
\caption{\label{G2r}$G_2(r)$ (defined in section \ref{Nematic}) as
a function of $r_{||}$ and $r_{\perp}$: The phases are very highly
ordered and only a detailed subsystem analysis confirmed the
quasi-long-ranged (rather than true-long-ranged) order at area
fraction $\phi$ = 60\% and 70\% and short-ranged correlations at
$\phi$ = 50\%. The decay exponents and correlation lengths
extracted from the plots are summarised in table \ref{exptab}.}
\end{center}
\end{figure}

As is evident from the real space images (Fig. \ref{realspace})
the long bails do indeed show nematic correlations. The tendency
to nematic order is quantified in Fig. \ref{G2r} which shows $G_2$
as a function of $r_{\perp}$ and $r_{||}$ for 3 different area
fractions $\phi$. Here $G_2(r) = <cos[2{(\theta_i - \theta_j)}]>$.
The symbols are the same as defined in section \ref{tetratic}.
The system shows a high degree of order at each of these area
fractions. The decay of $G_2$ was consistent with a power-law for
all three values of $\phi$, although the decay exponents (listed
in Table \ref{exptab}) were exceedingly small for the two higher
values of $\phi$.


\begin{table*}
\begin{center}
\caption{\label{exptab}The table below lists the correlation
lengths $\xi$ (scaled by $D$, the rod thickness) and decay
exponents $\alpha$ for, respectively, the short-range-ordered and
power-law-ordered nematic phases.}
\begin{tabular}{|c|cc|cc|}
\hline
  Area Fraction & $\xi(r_{||})/D$ & $\xi(r_{\perp})/D$ & $\alpha(r_{||})$  & $\alpha(r_{\perp})$ \\
  \hline
  50\% & 69 & 64 &  -  & - \\
  60\% & - & - & 0.0113 $\pm$ 4e-4 & 0.0182 $\pm$ 8e-4 \\
  70\% & - & - & 0.0066 $\pm$ 3e-4 & 0.0140 $\pm$ 5e-4 \\
\hline
\end{tabular}
\end{center}
\end{table*}

\begin{figure}
\begin{center}
\includegraphics[width = 3.25 in]{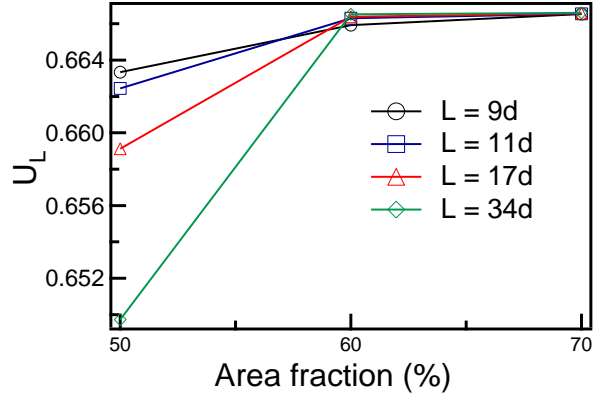}
\caption{\label{ULvsAF}Binder reduced cumulant $U_L$ vs area
fraction suggests quasi-long-range nematic order at and above
$\phi$ = 60\%, short-ranged order below. Here $L$ is the linear
size of the subsystem in units of the rod diameter $d$}
\end{center}
\end{figure}

Given the small system sizes in our experiment, we expect strong
finite size effects. In order to assess more precisely whether
the correlations were long-ranged, power-law, or short-ranged with
very long correlation lengths, we performed a subsystem analysis
of the cumulants following \cite{Binder}. The quantities evaluated
were $S = <\psi>_L$, where $\psi$ is the order parameter ($\psi =
\frac{1}{N_L}\sum e^{i2\theta_j}$, where $j$ runs over the number
$N_L$ of particles within the subsystem defined by $L$) and $L$ is
the linear size of the subsystem over which $S$ was evaluated,
$S^2 = <\psi ^2>_L$ and $S^4 = <\psi ^4>_L$. The Binder reduced
cumulant, $U_L = 1 - \frac{<S^4>}{3<S^2>^2}$ was then evaluated.
The results are plotted as a function of area fraction, $\phi$,
for different subsystem sizes $L$ in figure \ref{ULvsAF}.

\begin{figure}
\begin{center}
\includegraphics[width = 3.25 in]{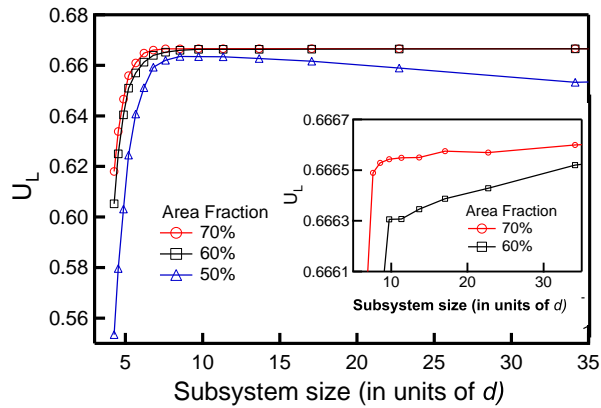}
\caption{\label{nematic_ULvsSS}Binder reduced cumulant $U_L$ vs
linear size $L$ of subsystem over which $U_L$ was evaluated. For
$\phi = 50$ \%, mild dip in curve suggests short-ranged order.
Inset: $U_L$ for $\phi$ = 60\% and 70\% at a higher resolution.
$\phi$ is very slightly, but systematically higher for $\phi$ =
70\%, leaving open the possibility of true long-range nematic
order in this driven two-dimensional systems.  Here $L$ is the
linear size of the subsystem in units of the rod diameter $d$}
\end{center}
\end{figure}

As shown in \cite{Binder}, for phases with LRO the curves for
different system sizes $L$ will cross at a certain value of area
fraction $\phi =\phi^*$. If there is QLRO, the curves of $U_L$ vs
$\phi$ for different $L$ should collapse onto a single line over
an extended range of values of $\phi$. The lower limit of this
value will be $\phi^*$, the isotropic-nematic transition value.
Figure \ref{ULvsAF} suggests that the phases formed at the higher
area fractions do show QLRO and the isotropic- nematic transition
occurs at an area fraction somewhere between 50\% and 60\%. In 
Fig.\ref{nematic_ULvsSS} we show the Binder cumulant
as a function of subsystem sizes $L$, for different area
fractions, $\phi$. The downturn at large $L$ for  $\phi=50\%$
indicates short-range order. We show a greatly magnified version
of this plot for the two higher area fractions as an inset. 
$U_L$ for 70\% is slightly but systematically larger than that
for 60\%. This difference is at the edge of our resolution, but
it leaves open the tempting possibility that these
nonequilibrium  nematics have true, rather than quasi- long-range
order. Further measurements on large systems are required to
settle this important question of principle.

\subsubsection{\label{Amp}Varying amplitude of oscillation}

The presence of a third dimension into which the particle is
lofted means that the 2-dimensional projections of the particles
can overlap and that unlike hard particles, a temperature variable
could be relevant to the order.  As we said in subsection
\ref{Cell}, the shaking amplitude $a_{o}$ could act as a kind of
temperature. A natural dimensionless measure of how strongly the
system is being shaken is $\Gamma \equiv (2 \pi f)^2 a_{o}/g$,
where $f$ and $g$ are, respectively, the oscillation frequency and
the acceleration due to gravity.  We studied the effect on the
nematic order of changing acceleration while holding fixed the gap
height, $H$ and particle diameter, $D$; changing these would have
changed the inter-penetrability of the particles in two
dimensions.

\begin{figure}
\begin{center}
\includegraphics[width = 3.25 in]{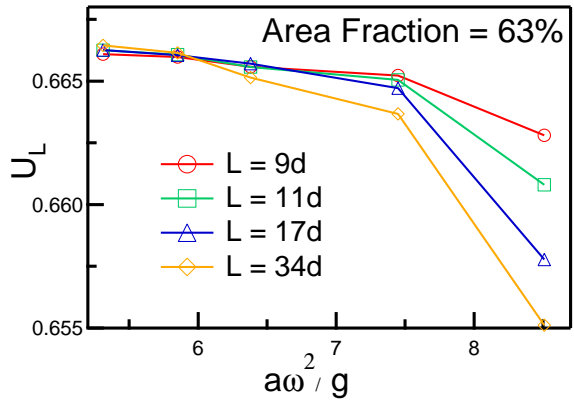}
\caption{\label{diff_amplitudes}Binder reduced cumulant $U_L$
versus dimensionless acceleration. The loss of long-ranged order
upon increasing amplitude confirms the notion that a temperature
variable is relevant in this system and is related to the shaking
strength.}
\end{center}
\end{figure}

Figure \ref{diff_amplitudes} shows the Binder reduced cumulant of
the phase as a function of dimensionless acceleration. The area
fraction of the phase is about 63\%, within the nematic regime
identified in Fig.\ref{nematic_ULvsSS}. The acceleration is varied
by holding the frequency, $f$, fixed and varying the amplitude,
$a_{o}$. Even though $a_{o}<< (H-D)$ over the entire range of
amplitudes, the higher shaking amplitudes do affect the extent of
order in the system. This is not entirely surprising: while
$a_{o}$ is small, the ratio of the characteristic kinetic energy
$1/2m(2 \pi fa_{o})^2$ and the maximum possible potential energy
$mg(H-D)$ varies from $0.2$ to $0.36$ over the range of the plot.
However, as can be seen by the small range of vertical axis, the
degree to which the phase disorders by increasing $\Gamma$ from 5
to 8 is much less than that achieved by reducing the area fraction
from 70\% to 50\% .

\subsection{\label{smectics}Two-dimensional smectic order?}

Our observations on basmati rice gave images with strong
indications of striped, i.e., smectic order (see Fig.
\ref{realspace}). In two-dimensional systems at thermal
equilibrium, we know that thermal fluctuations render smectic
order strictly short-ranged, while in three-dimensional systems it
is quasi-long-ranged. These conclusions are based on derivations
of the variance of layer displacement fluctuations from an elastic
free-energy functional, in which, as a consequence of
rotation-invariance, the cost of layer displacements with
wavevector ${\bf q}$ parallel to the layers varies as $q^4$. Since
the systems we are studying are driven, not thermal equilibrium,
these arguments do not apply {\it a priori}, and the possibility
of quasi-long-range (or, for that matter, true long-range) smectic
order in our vibrated granular- rod monolayers remains open. We
are studying this possibility theoretically \cite{usunpb}.
Experimentally, however, our systems are not large enough for the
type of analysis (see, e.g., \cite{caille,safinya}) required to
check whether the images we are seeing are true smectics, weakly
ordered two-dimensional crystals, or smectics disordered by the
presence a finite but small density of free dislocations (which
would be nematics on large length scales).

\subsection{\label{Noneqphen}Dynamical nonequilibrium
phenomena}

\subsubsection{\label{Rotation}Collective, coherent rotation}

All the phases showed a marked tendency to rotate globally. For
the nematic phases, the square frame prevented them from rotating 
(except in certain cases, which we shall come to in the next
paragraph), but we could not prevent the tetratic phase from
rotating, in part because we did not use a square frame, since we
wanted to see how strong the tetratic correlations were in the
absence of such externally imposed four-fold alignment.

Since this is a nonequilibrium system, any stray chirality in the
system, even at the boundary, should result \cite{curieprincip} in
macroscopic circulation. The observed global rotation is thus a
signature of the driven nature of the system. A thermal
equilibrium system, no matter how chirally asymmetric, would not
display persistent currents. Our observed rotation rates were
of the order of a degree in $10^4$ cycles of oscillation. In one
instance, when we were using the bails in the square boundary
(where, under normal circumstances, the rotations are curbed), the
phase began rotating fast, turning through a degree in $\approx
10^3$ cycles of oscillation. The source of the asymmetry appeared
to be a slight misalignment of the particles fixed at the
boundary, because we were able to stop the rotation by aligning
them more carefully. That this minor imperfection in boundary
alignment could have such a strong effect was a surprise, and
means the rotation phenomenon merits more serious investigation.

\subsubsection{\label{Swirls}Swirls}

Since \textit{basmati} rice (aspect ratio ${\cal A}= 4.2$) showed
smectic tendencies, we studied tapered copper rods with similar
aspect ratio (aspect ratio ${\cal A}= 4.9$) but half the length of
the rice grains, in the hope of getting a larger number of smectic
layers in the sample cell. Unexpectedly, we found no sign of
smectic layering. Instead, a very dynamic pattern of swirls
appeared, reminiscent of bacterial swimming patterns
\cite{dombrowski}. The swirly state looks like a defect- ridden
nematic, which is curious because equilibrium Monte Carlo studies
on rods in two dimensions suggest that nematics should not form
for aspect ratios less than 7. These swirls, which are strength
$\pm 1/2$ disclinations, differ from those seen in \cite{blair}:
the latter are strength 1, they have a core that looks like a
crystalline packing of standing rods, and each defect is chiral
because the rods have a systematic tilt in the direction of their
tangential velocity orientation of the rods in the region outside
the core. In our case, the circulation appears to be a result of
the interaction between defects. It remains to be seen whether we
are seeing coarsening towards a nematic, the disordering of a
nematic (i.e., coarsening towards the isotropic phase), or simply
a disordered, correlated, defect-turbulent steady state.

\section{\label{conclusion}Conclusion and Discussion}

This paper has presented an exploration of the collective
properties of vibrated rods in two dimensions, focusing on
orientationally ordered phases. Tapered particles lead to nematic
order at large area fractions, in keeping with the thermal
equilibrium Monte Carlo studies of \cite{Bates_Frenkel}, while
strictly cylindrical particles, with flat tips, display strong
tetratic correlations over a very broad range of aspect ratios. It
remains to be established whether the strong sensitivity to
particle shape is a nonequilibrium phenomenon or whether it can be
reproduced in equilibrium simulations. We also observed some
clearly nonequilibrium phenomena: persistent, systematic rotation
of the entire phase, presumably as a result of stray chirality in
the system, and also a rich pattern of circulating disclinations
of strength $\pm$ 1/2. Detailed quantitative analyses of the
dynamics of the swirls, of tagged-particle motion and
time-dependent correlations of collective modes and other
nonequilibrium effects will appear in a separate paper
\cite{usunpb}.

\section{\label{Ack!!!}Acknowledgements}
We thank P.R. Nott and V. Kumaran, Dept. of Chemical Engineering,
IISc, and A.K. Raychaudhuri, Dept of Physics, IISc, for generously
letting us use their experimental facilities. We thank S. Fraden
for detailed and helpful comments and for sharing his unpublished
results with us. We thank M. Barma, M. Khandkar, and  D. Dhar for
very valuable discussions. VN thanks Sohini Kar for help with some
of the experiments. The work at UMass Amherst was supported by
funds from NSF-DMR 0305396. The Centre for Condensed Matter Theory
is supported by the Department of Science and Technology, India.
VN and NM thank, respectively, the Department of Physics,
University of Massachusetts, Amherst and the Department of
Physics, Indian Institute of Science, for support and hospitality
while part of this work was done. VN thanks the Council for
Scientific and Industrial Research, India and the Indian Institute
of Science for providing travel aid.


\begin{thebibliography}{10}
\bibitem{bailfootnote}
The particles we call ``bails'' resemble a piece of cricket
equipment of the same name.
\end{thebibliography}

\section{\label{References}References}

\end{document}